\documentclass[a4paper,10pt]{article}  
\usepackage[]{graphicx}
\usepackage[]{amsmath}
\usepackage[]{amssymb}
\usepackage[]{bm}
\usepackage[]{slashed}
\title{Quest for MeV frequency combs -- proposal for ELI experiments\footnote{Based on the talk presented at the \textsl{19$^{\mathrm{th}}$ Polish-Slovak-Czech Optical Conference on Wave and Quantum Aspects of Contemporary Optics}, September 8--12, 2014, Wojan\'ow Palace, Poland}
} 

\author{Katarzyna Krajewska and Jerzy Z. Kami\'{n}ski\footnote{E-mail: Jerzy.Kaminski@fuw.edu.pl} \\
Institute of Theoretical Physics, Faculty of Physics, \\ University of Warsaw,
Pasteura~5,~02-093~Warszawa,~Poland
}

\date{\empty}

\begin{document} 
  \maketitle 

%%%%%%%%%%%%%%%%%%%%%%%%%%%%%%%%%%%%%%%%%%%%%%%%%%%%%%%%%%%%% 
\begin{abstract}
The optical frequency comb has become an indispensable tool for high precision spectroscopy. Also experiments in the field of ultrafast physics rely on the frequency comb technique to generate precisely controlled attosecond optical pulses by means of the high-order harmonic generation. However, in order to generate even shorter laser pulses or to apply this technique in investigations of nuclear structure, combs of frequencies of the order of MeV are necessary. It seems that it may not be possible to achieve such photon energies by high-order harmonic generation. In this context the possibility of the generation of Thomson and Compton-based frequency combs is presented. Diffraction of generated radiation by a sequence of laser pulses and its analogy to the diffraction grating is elucidated. Theoretical investigations presented in this report can be considered as the proposal for future ELI experiments [www.eli-laser.eu]
\end{abstract}

%\keywords{Frequency combs, high-order harmonics, zeptosecond pulses}

%%%%%%%%%%%%%%%%%%%%%%%%%%%%%%%%%%%%%%%%%%%%%%%%%%%%%%%%%%%%%
\section{INTRODUCTION}
\label{sec:intro}  

Historically, strong-field classical and  quantum electrodynamics are the oldest strong-field-physics areas, as investigations of fundamental electromagnetic processes in laser fields began already in the 1960s (for recent reviews see, e.g., Refs. 1--4). At that time, the problem was of purely theoretical interest and the main focus was on a qualitative understanding of effects imposed by intense and mostly monochromatic electromagnetic radiation. With the rapid development of high-power laser technology, that we have encountered over the last decades, the laser radiation in the near-visible spectrum with intensities as high as $10^{22}$W/cm$^2$ can be produced in the laboratory. At the same time, the synthesis of laser pulses with well-controlled properties, such as the carrier-envelope phase or the envelope shape, is currently feasible. Due to those recent advances, it became urgent to quantitatively investigate more subtle effects of the laser-field interaction with matter than was originally pursued. The aim of this presentation is to discuss some selected problems related to generation of high-frequency radiation by interaction of relativistic electrons with intense laser pulses. In particular we analyze the high-order harmonic generation, formation of the coherent comb structures in the MeV domain with possible applications in nuclear physics \cite{Palffy2013,Palffy2014}, and synthesis of zepto- or even yoctosecond pulses \cite{Liu2012,Krajewska2014b}.

The `classical' Young double-slit experiment \cite{Young1804} is still considered as a trailblazer for modern explorations of quantum phenomena. Not surprisingly, the ideas that are behind this experiment are currently vigorously investigated not only in quantum optics and information theory, but also in the strong-field classical and quantum electrodynamics. As an example, we mention the theoretical proposal \cite{King2010} of the so-called `matterless double-slit experiment' or the Kapitza-Dirac effect \cite{Kapitza1933,Fedorov1991,Fedorov1997,Ahrens2012}, i.e., the diffraction of electrons by a standing electromagnetic wave. The aim of this report is to present recently investigated diffraction of free electrons by modulated finite laser pulses \cite{Krajewska2014a,Krajewska2014c}, that shows very similar features observed for the diffraction grating, but in the frequency domain. The diffraction pattern is going to be analyzed within the nonlinear Thomson and Compton scattering, that are straightforward generalizations of their weak-field analogues discovered by J.~J.~Thomson \cite{Thomson1906} for the classical electrodynamics and by A.~H.~Compton \cite{Compton1923a,Compton1923b} for the quantum electrodynamics.

The organization of this report is the following. In Secs.~\ref{Thomson} and \ref{Compton} we present the theoretical formulations for the electron scattered by a laser pulse with the emission of electromagnetic radiation within the classical and quantum electrodynamics, respectively. Then, in Sec.~\ref{Diffraction} we derive the diffraction formulas for both theories and illustrate them for particular cases. The pulse synthesis from the frequency distribution is discussed in Sec.~\ref{Pulse}, and in Sec.~\ref{Conclusions} we draw some concluding remarks.

Throughout the paper, we keep $\hbar=1$. Hence, the fine-structure constant equals $\alpha=e^2/(4\pi\varepsilon_0c)$. We use this constant in expressions derived from classical electrodynamics as well, where it is meant to be 
multiplied by $\hbar$ in order to restore the physical units. In numerical analysis we use relativistic units (rel. units) such that $\hbar=m_{\rm e}=c=1$ where $m_{\rm e}$ is the electron rest mass. Furthermore, we denote the product of any two four-vectors $a^{\mu}$ and $b^{\mu}$ with $a\cdot b = a^{\mu}b_{\mu}=a^0b^0-a^1b^1-a^2b^2-a^3b^3$ ($\mu = 0,1,2,3$), where the Einstein summation convention is used. We employ the Feynman notation $\slashed{a} = \gamma\cdot a=\gamma^{\mu} a_{\mu}$ for the contraction with the Dirac matrices $\gamma^{\mu}$ and use a customary notation $\bar{u}=u^{\dagger}\gamma^0$, where $u^{\dagger}$ is the Hermitian conjugate of $u$. Finally, we use the so-called light-cone variables. Namely, for a given space direction determined by a unit vector $\bm{n}$ (which in our presentation is the direction of the laser pulse propagation) and for an arbitrary four-vector $a$, we keep the following notations: $a^{\|}=\bm{n}\cdot\bm{a}$, $a^-=a^0-a^{\|}$, $a^+=(a^0+a^{\|})/2$, and $\bm{a}^{\bot}=\bm{a}-a^{\|}\bm{n}$. For the four-vectors we use both the contravariant $(a^0,a^1,a^2,a^3)$ and the standard $(a_0,a_x,a_y,a_z)=(a_0,\bm{a})$ notations.

\section{Thomson Scattering}
\label{Thomson}

The complete theory of the Thomson scattering is presented in the textbooks \cite{Jackson1975,Landau1987}. Therefore, the aim of this section is to settle the notation that is used further in this report.
Our aim is to derive the frequency-angular distribution of the electromagnetic energy that is emitted during either Compton or Thomson scattering in the form of outgoing spherical waves. Their polarization is given by a complex unit vector $\bm{\varepsilon}_{\bm{K}\sigma}$, where $\sigma=\pm$ labels two polarization degrees of freedom, and where $\bm{K}$ is the wave vector of  radiation emitted in the direction $\bm{n}_{\bm{K}}$. Note that ${\bm K}$ determines also the frequency of the emitted radiation since $\omega_{\bm{K}}=c|\bm{K}|$. The wave four-vector, $K$, is therefore $K=(\omega_{\bm{K}}/c)(1,\bm{n}_{\bm{K}})$ where $K^2=0$ and $K\cdot\varepsilon_{\bm{K}\sigma}=0$ (we keep $\varepsilon_{\bm{K}\sigma}=(0,\bm{\varepsilon}_{\bm{K}\sigma})$ 
and $\varepsilon_{\bm{K}\sigma}\cdot\varepsilon_{\bm{K}\sigma'}^*=-\delta_{\sigma\sigma'}$). We also assume that three vectors, $(\bm{\varepsilon}_{\bm{K}+},\bm{\varepsilon}_{\bm{K}-},\bm{n}_{\bm{K}})$ 
form the right-handed system of mutually orthogonal unit vectors such that $\bm{\varepsilon}_{\bm{K}+}\times\bm{\varepsilon}_{\bm{K}-}=\bm{n}_{\bm{K}}$.

The laser field which drives both the nonlinear Compton and Thomson scattering is modeled as 
a linearly polarized, pulsed plane wave field, with the following vector potential,
\begin{equation}
\bm{A}(\phi)=A_0 N_{\mathrm{osc}}\bm{\varepsilon}f(\phi).
\label{t1}
\end{equation}
Here, the real vector $\bm{\varepsilon}$ determines the linear polarization of the pulse, and the shape function $f(\phi)$ is defined via its derivative
\begin{equation}
f'(\phi)\sim \begin{cases} 0, & \phi <0, \cr
                 \sin^2\bigl(N_{\mathrm{rep}}\frac{\phi}{2}\bigr)\sin(N_{\mathrm{rep}}N_{\mathrm{osc}}\phi), & 0\leqslant\phi\leqslant 2\pi,\cr
								0, & \phi > 2\pi,
			  \end{cases}
\label{t2}
\end{equation}
where we assume that $f(0)=0$. Above, $N_{\mathrm{osc}}$ and $N_{\mathrm{rep}}$ determine the number of cycles in the subpulse and the number of such subpulses, respectively. The function $f'(\phi)$ is normalized such that
\begin{equation}
\int_0^{2\pi}\mathrm{d}\phi\, [f'(\phi)]^2=\pi.
\end{equation}

\begin{figure}
   \begin{center}
   \begin{tabular}{c}
   \includegraphics[width=10cm]{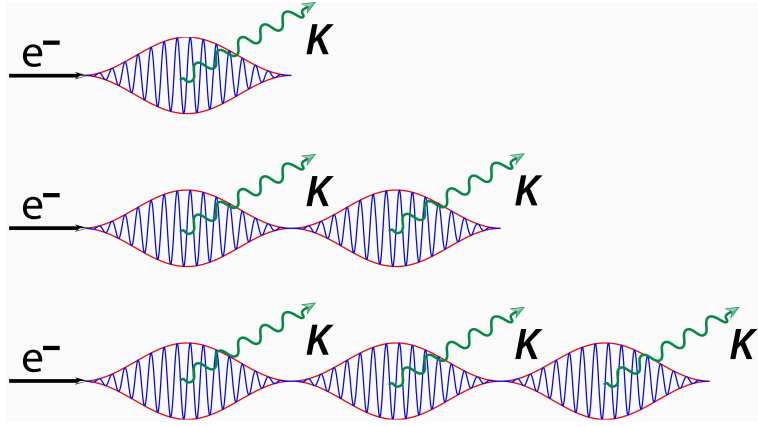}
   \end{tabular}
   \end{center}
   \caption
		{ \label{youngmoreplow} 
Young-type interference in the frequency domain for 1, 2 and 3 laser-field subpulses. Each subpulse corresponds to a particular slit in the original Young experiment. Since it is not known from which subpulse the Compton photon is emitted, therefore, the total probability amplitude shows the interference pattern, which appears as the frequency comb of generated radiation. For the Thomson scattering each subpulse emits its own spherical electromagnetic wave the interference of which leads to the formation of the comb.
   }
   \end{figure}

Let us further assume that the duration of the laser pulse is $T_{\mathrm{p}}$. This allows us to introduce 
the fundamental, $\omega=2\pi/T_{\mathrm{p}}$, and the central, $\omega_{\mathrm{L}}=N_{\mathrm{osc}}\omega$, frequencies
of the laser field. Moreover, if the laser pulse propagates in the direction given by the unit vector $\bm{n}$, 
we can define the laser-field four-vector $k=(\omega/c)(1,\bm{n})$ such that $k^2=0$. Hence, the phase $\phi$ in Eq.~\eqref{t1} becomes
\begin{equation}
\phi=k\cdot x=\omega\Bigl(t-\frac{\bm{n}\cdot\bm{r}}{c}\Bigr).
\label{t4}
\end{equation}
For our further purposes we introduce the dimensionless and relativistically invariant parameter
\begin{equation}
\mu=\frac{|e|A_0}{m_{\mathrm{e}}c},
\label{t5}
\end{equation}
where $e=-|e|$ is the electron charge. With these notations, the laser electric and magnetic fields are equal to
\begin{equation}
\bm{\mathcal{E}}(\phi)=\frac{\omega m_{\mathrm{e}}c\mu}{e}N_{\mathrm{osc}}\bm{\varepsilon}f'(\phi),\quad
\bm{\mathcal{B}}(\phi)=\frac{\omega m_{\mathrm{e}}\mu}{e}N_{\mathrm{osc}}(\bm{n}\times\bm{\varepsilon})f'(\phi).
\label{t6}
\end{equation}
Such pulses for $N_{\mathrm{osc}}=16$ and $N_{\mathrm{rep}}=1,2,3$ are illustrated in Fig.~\ref{youngmoreplow}.

%-------------
%-------------

The frequency-angular distribution of energy emitted during the Thomson process can be expressed as
\begin{equation}
\frac{\mathrm{d}^3E_{\mathrm{Th}}(\bm{K},\sigma)}{\mathrm{d}\omega_{\bm{K}}\mathrm{d}^2\Omega_{\bm{K}}}=\alpha|\mathcal{A}_{\mathrm{Th},\sigma}(\omega_{\bm{K}})|^2.
\label{t11}
\end{equation}
The complex function $\mathcal{A}_{\mathrm{Th},\sigma}(\omega_{\bm{K}})$ will be called here the Thomson amplitude and its explicit form can be represented as an integral 
\begin{equation}
\mathcal{A}_{\mathrm{Th},\sigma}(\omega_{\bm{K}})=\frac{1}{2\pi}\int_0^{2\pi}\mathrm{d}\phi \Upsilon_{\sigma}(\phi)\exp(\mathrm{i}\omega_{\bm{K}}\ell(\phi)/c),
\label{t12}
\end{equation}
where
\begin{equation}
\ell(\phi)=c\frac{\phi}{\omega}+(\bm{n}-\bm{n}_{\bm{K}})\cdot \bm{r}(\phi),\quad
\Upsilon_{\sigma}(\phi)=\bm{\varepsilon}_{\bm{K}\sigma}^*\cdot \frac{\bm{n}_{\bm{K}}\times [(\bm{n}_{\bm{K}}-\bm{\beta}(\phi))\times \bm{\beta}'(\phi)]}{\bigl(1-\bm{n}_{\bm{K}}\cdot \bm{\beta}(\phi)\bigr)^2},
\end{equation}
and where the electron position $\bm{r}(\phi)$ and reduced velocity $\bm{\beta}(\phi)$ fulfill the system of ordinary differential equations
\begin{equation}
\frac{\mathrm{d}\bm{r}(\phi)}{\mathrm{d}\phi}=\frac{c}{\omega}\frac{\bm{\beta}(\phi)}{1-\bm{n}\cdot\bm{\beta}(\phi)}, \quad
\frac{\mathrm{d}\bm{\beta}(\phi)}{\mathrm{d}\phi}=\mu\frac{\sqrt{1-\bm{\beta}^2(\phi)}}{1-\bm{n}\cdot\bm{\beta}(\phi)} \Bigl[ 
\bigl(\bm{\varepsilon}-\bm{\beta}(\phi)(\bm{\beta}(\phi)\cdot\bm{\varepsilon})+\bm{\beta}(\phi)\times(\bm{n}\times\bm{\varepsilon})\bigr)f^{\prime}(\phi) \Bigr] . \label{thom9ex}
\end{equation}
These equations can be derived from the Newton-Lorentz relativistic equations, with time $t$ which relates to the phase $\phi$ by Eq.~\eqref{t4}.

\section{Compton Scattering}
\label{Compton}

The probability amplitude for the Compton process, $e^-_{\bm{p}_{\mathrm{i}}\lambda_{\mathrm{i}}}\rightarrow e^-_{\bm{p}_{\mathrm{f}}\lambda_{\mathrm{f}}}+\gamma_{\bm{K}\sigma}$, with the initial and final electron momenta and spin polarizations $\bm{p}_{\mathrm{i}}\lambda_{\mathrm{i}}$ and $\bm{p}_{\mathrm{f}}\lambda_{\mathrm{f}}$, 
respectively, equals
\begin{equation}
{\cal A}(e^-_{\bm{p}_{\mathrm{i}}\lambda_{\mathrm{i}}}\rightarrow e^-_{\bm{p}_{\mathrm{f}}\lambda_{\mathrm{f}}}
+\gamma_{\bm{K}\sigma})=-\mathrm{i}e\int \mathrm{d}^{4}{x}\, j^{(++)}_{\bm{p}_{\mathrm{f}}\lambda_{\mathrm{f}},
\bm{p}_{\mathrm{i}}\lambda_{\mathrm{i}}}(x)\cdot A^{(-)}_{\bm{K}\sigma}(x), \label{ComptonAmplitude}
\end{equation}
where $\bm{K}\sigma$ denotes the Compton photon momentum and polarization. Here, we consider the case when both the laser pulse and the Compton photon are linearly polarized. In Eq.~\eqref{ComptonAmplitude},
\begin{equation}
A^{(-)}_{\bm{K}\sigma}(x)=\sqrt{\frac{1}{2\varepsilon_0\omega_{\bm{K}}V}} 
\,\varepsilon_{\bm{K}\sigma}\mathrm{e}^{\mathrm{i}K\cdot x},
\label{per}
\end{equation}
where $V$ is the quantization volume and $j^{(++)}_{\bm{p}_{\mathrm{f}} \lambda_{\mathrm{f}},\bm{p}_{\mathrm{i}}\lambda_{\mathrm{i}}}(x)$ is the matrix element of the electron current operator with its $\nu$-component equal to
\begin{equation}
[j^{(++)}_{\bm{p}_{\mathrm{f}} \lambda_{\mathrm{f}},\bm{p}_{\mathrm{i}}\lambda_{\mathrm{i}}}(x)]^{\nu}
=\bar{\psi}^{(+)}_{\bm{p}_{\mathrm{f}} \lambda_{\mathrm{f}}}(x)\gamma^\nu \psi^{(+)}_{\bm{p}_{\mathrm{i}}\lambda_{\mathrm{i}}}(x).
\end{equation}
Here, $\psi^{(+)}_{\bm{p}\lambda}(x)$ is the so-called Volkov solution of the Dirac equation coupled to the electromagnetic field,
\begin{equation}
\psi^{(+)}_{\bm{p}\lambda}(x)=\sqrt{\frac{m_{\mathrm{e}}c^2}{VE_{\bm{p}}}}\Bigl(1-\frac{e}{2k\cdot p}\slashed{A}\slashed{k}\Bigr)
u^{(+)}_{\bm{p}\lambda}\mathrm{e}^{-\mathrm{i} S_p^{(+)}(x)} , \quad
S_p^{(+)}(x)=p\cdot x+\int^{k\cdot x} \Bigl[\frac{ e A(\phi )\cdot p}{k\cdot p}
-\frac{e^2A^{2}(\phi )}{2k\cdot p}\Bigr]{\rm d}\phi ,
\label{Volk}
\end{equation}
where $E_{\bm{p}}=cp^0$, $p=(p^0,\bm{p})$, $p\cdot p=m_{\mathrm{e}}^2c^2$, and $u^{(+)}_{\bm{p}\lambda}$ 
is the free-electron bispinor normalized such that $\bar{u}^{(+)}_{\bm{p}\lambda}u^{(+)}_{\bm{p}\lambda'}=\delta_{\lambda\lambda'}$. The four-vector potential $A(k\cdot x)$ in Eq.~\eqref{Volk} represents an external electromagnetic radiation generated by lasers, in the case when a transverse variation of the laser field in a focus is negligible. In other words, $A(k\cdot x)$ represents the plane-wave-fronted pulse. In this case, $k\cdot A(k\cdot x)=0$ and $k\cdot k=0$, which allows one to exactly solve the Dirac equation for such electromagnetic fields. 

%-------------
   \begin{figure}
   \begin{center}
   \begin{tabular}{c}
   \includegraphics[width=11.5cm]{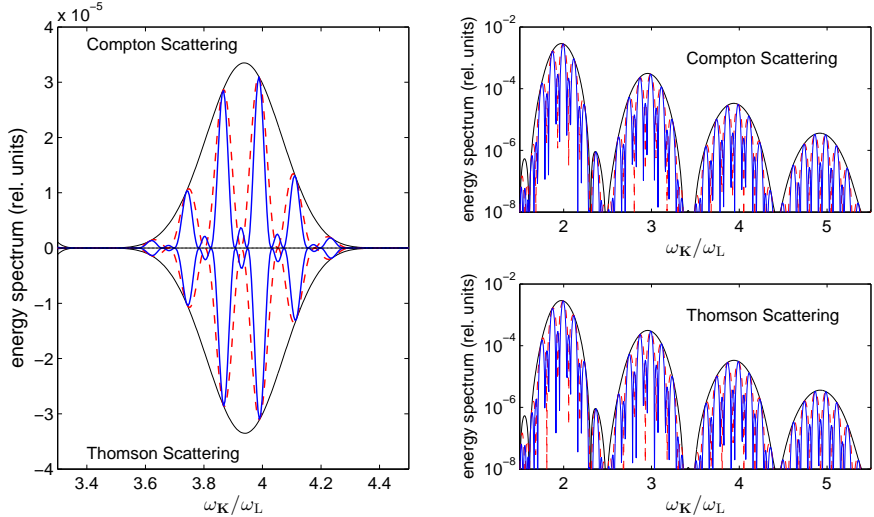}
   \end{tabular}
   \end{center}
   \caption
		{ \label{tcomb1} 
		Compton and Thomson energy distributions as a functions of frequency $\omega_{\bm{K}}$. The laser beam, linearly polarized in the $x$-direction, propagates in the $z$-direction and collides with the electron beam in the head-on geometry. The distribution is calculated in the reference frame of electrons with the laser pulse parameters such that $\omega_{\mathrm{L}}=3\times 10^{-3}m_{\mathrm{e}}c^2$, $\mu=1$, $N_{\mathrm{osc}}=8$. The emitted radiation is calculated for $\theta_{\bm{K}}=0.1\pi$ and $\varphi_{\bm{K}}=0$. The thin black line (the envelope) corresponds to $N_{\mathrm{rep}}=1$, the thick dashed red line to $N_{\mathrm{rep}}=2$, the thick blue line to $N_{\mathrm{rep}}=3$, and the distributions are divided by $N_{\mathrm{rep}}^2$. In the left frame we present the Compton and Thomson (reflected with respect to the horizontal black line) distributions on the linear scale. The two frames on the right-hand side show the same distributions on the logarithmic scale and for the larger frequency domain. For the geometry considered and the laser pulse parameters these two distributions are nearly identical. 
   }
   \end{figure} 
%-------------

The probability amplitude for the Compton process~\eqref{ComptonAmplitude} becomes
\begin{equation}
{\cal A}(e^-_{\bm{p}_\mathrm{i}\lambda_\mathrm{i}}\longrightarrow e^-_{\bm{p}_\mathrm{f}\lambda_\mathrm{f}}+\gamma_{\bm{K}\sigma})
=\mathrm{i}\sqrt{\frac{2\pi\alpha c(m_{\mathrm{e}}c^2)^2}{E_{\bm{p}_\mathrm{f}}E_{\bm{p}_\mathrm{i}}\omega_{\bm{K}}V^3}}\, \mathcal{A}, \label{ct1}
\end{equation}
where 
\begin{equation}
\mathcal{A}= \int\mathrm{d}^{4}{x}\,   \bar{u}^{(+)}_{\bm{p}_\mathrm{f}\lambda_\mathrm{f}}\Bigl(1-\mu\frac{m_\mathrm{e}c}{2p_\mathrm{f}\cdot k}f(k\cdot x)\slashed{\varepsilon}\slashed{k}\Bigr)\slashed{\varepsilon}_{\bm{K}\sigma}  \,
\Bigl(1+\mu\frac{m_\mathrm{e}c}{2p_\mathrm{i}\cdot k} f(k\cdot x)\slashed{\varepsilon}\slashed{k}\Bigr) u^{(+)}_{\bm{p}_\mathrm{i}\lambda_\mathrm{i}}\,\mathrm{e}^{-\mathrm{i}S(x)},  \label{ct2}
\end{equation}
with $S(x) = S^{(+)}_{{p}_\mathrm{i}}(x)-S^{(+)}_{{p}_\mathrm{f}}(x)-K\cdot x$. While moving in a laser pulse, the electron acquires an additional momentum shift~\cite{Krajewska2012}. This leads  to a notion of the laser-dressed momentum:
\begin{equation}
\bar{p} = p-\mu m_{\rm e}c\frac{p\cdot\varepsilon}{p\cdot k}\left\langle f\right\rangle k 
+\frac{1}{2}(\mu m_{\rm e}c)^2\frac{1}{p\cdot k}\left\langle f^2\right\rangle k\,.
\end{equation}
The gauge-invariant form of dressing is discussed in Ref.~23. Having this in mind we can define
\begin{equation}
N_{\rm eff} = \frac{K^{0}+\bar{p}^{0}_{\rm f}-\bar{p}^{0}_{\rm i}}{k^{0}}=
cT_{\rm p}\frac{K^{0}+\bar{p}^{0}_{\rm f}-\bar{p}^{0}_{\rm i}}{2\pi},
\end{equation}
which is both gauge- and relativistically invariant~\cite{Krajewska2012}.

The frequency-angular distribution of energy of the emitted photons for an unpolarized electron beam is given by
\begin{equation}\label{copton:spectrum:3}
\frac{{\rm d^{3}}E_{\rm C}}{{\rm d}\omega_{\bm K}{\rm d^{2}}\Omega_{\bm K}}
=\frac{1}{2}\sum_{\sigma=1,2}\sum_{\lambda_{\rm i}=\pm}\sum_{\lambda_{\rm f}=\pm}
\frac{{\rm d^{3}}E_{{\rm C},\sigma}(\lambda_{\rm i},\lambda_{\rm f})}{{\rm d}\omega_{\bm K}{\rm d^{2}}\Omega_{\bm K}},
\end{equation}
where
\begin{equation}\label{copton:spectrum:2}
\frac{{\rm d^{3}}E_{{\rm C},\sigma}(\lambda_{\rm i},\lambda_{\rm f})}{{\rm d}\omega_{\bm K}{\rm d^{2}}\Omega_{\bm K}}
=\alpha\left|\mathcal{A}_{{\rm C},\sigma}(\omega_{\bm{K}},\lambda_{\rm i},\lambda_{\rm f})\right|^2,
\end{equation}
and the scattering amplitude equals
\begin{equation}\label{copton:spectrum:1}
\mathcal{A}_{{\rm C},\sigma}(\omega_{\bm{K}},\lambda_{\rm i},\lambda_{\rm f}) 
=\frac{m_{\rm e}c K^{0}}{2\pi\sqrt{p_{\rm i}^{0}k^{0}(k\cdot p_{\rm f})}}
\sum_{N}D_{N}\frac{1-{\rm e}^{-2\pi\mathrm{i}(N-N_{\rm eff})}}{\mathrm{i}(N-N_{\rm eff})},
\end{equation}
with the functions $D_N$ defined in Ref. 22. %~\cite{Krajewska2012}.

In Fig.~\ref{tcomb1} we compare the classical Thomson process with its quantum analogue, the Compton one, for laser pulses consisting of up to three subpulses, as schematically presented in Fig.~\ref{youngmoreplow}. The presentation is for relatively long subpulses, consisting of $N_{\mathrm{osc}}=8$ laser field oscillations. In this case, we obtain for $N_{\mathrm{rep}}=1$ for both the classical and quantum cases the energy distribution consisting of the well-separated from each other and broad peaks, which does not resemble the frequency comb. However, for $N_{\mathrm{rep}}>1$ the sharp peaks appear. They tend to become more narrow with increasing $N_{\mathrm{rep}}$, but they appear for the same frequencies independent of $N_{\mathrm{rep}}$. Moreover, the height of the individual peak scales as $N_{\mathrm{rep}}^2$, which indicates the coherence of the generated comb. The numerical analysis of the phase of the Compton and Thomson amplitudes shows that, at the peak frequencies, phases are equal to 0 modulo $\pi$. In addition, the derivatives of these phases with respect to $\omega_{\bm{K}}$ are almost constant (in the considered domain of $\omega_{\bm{K}}$). This proves that the separation between the consecutive peaks is nearly the same; hence, a coherent and equally spaced frequency comb is created. In order to analyze theoretically the properties of the frequency combs we present below the diffraction formulas for the Thomson and Compton amplitudes.

\section{Diffraction Formulas}
\label{Diffraction}

In this section we discuss the diffraction formulas for the Thomson and Compton amplitudes that prove the 'phase-matching' conditions for the peaks in the energy distributions at which the global phases change by $\pi$. We also show that, although for classical theory this can happen for the equally spaced frequencies, for quantum theory this is not exactly the case.

In classical electrodynamics, by applying the symmetry properties of the modulated laser pulse considered in this report, one can show that the Thomson amplitude adopts the following diffraction-type form \cite{Krajewska2014c}
\begin{align}
 \mathcal{A}_{\mathrm{Th},\sigma}(\omega_{\bm{K}})=&\exp\Bigl[\mathrm{i}\Phi_{\mathrm{Th},\sigma}(\omega_{\bm{K}})\Bigr] \frac{\sin\Bigl[\frac{\omega_{\bm{K}}N_{\mathrm{rep}}}{2c}\ell\Bigl(\frac{2\pi}{N_{\mathrm{rep}}}\Bigr)\Bigr]}
 {\sin\Bigl[\frac{\omega_{\bm{K}}}{2c}\ell\Bigl(\frac{2\pi}{N_{\mathrm{rep}}}\Bigr) \Bigr]}|\mathcal{A}^{(1)}_{\mathrm{Th},\sigma}(\omega_{\bm{K}})|,
\label{bi4}
\end{align}
where $|\mathcal{A}^{(1)}_{\mathrm{Th},\sigma}(\omega_{\bm{K}})|$ is the absolute value of the Thomson amplitude for the single subpulse and the Thomson global phase equals
\begin{equation}
\Phi_{\mathrm{Th},\sigma}(\omega_{\bm{K}})=\Bigl(N_{\mathrm{rep}}\mp\frac{1}{2}\Bigr)\pi+
N_{\mathrm{rep}}\frac{\omega_{\bm{K}}}{c} \ell\Bigl(\frac{\pi}{N_{\mathrm{rep}}}\Bigr).
\label{bi13}
\end{equation}
For particular frequencies $\omega_{\bm{K},L}$ that fulfill the condition
\begin{equation}
 \frac{\omega_{\bm{K},L}}{c}\ell\Bigl(\frac{2\pi}{N_{\mathrm{rep}}}\Bigr)=2\pi L,\quad L=1,2,\dots\, ,
\label{bi6}
\end{equation}
we have the diffraction enhancement of the energy distribution generated by Thomson scattering 
(similar to the diffraction grating pattern for the angular distribution), as $|\mathcal{A}_{\mathrm{Th},\sigma}(\omega_{\bm{K},L})|^2=N_{\mathrm{rep}}^2 |\mathcal{A}^{(1)}_{\mathrm{Th},\sigma}(\omega_{\bm{K},L})|^2$. Moreover, for $N_{\mathrm{rep}}>1$, the Thomson amplitude vanish for $\omega_{\bm{K}}$ such that
\begin{equation}
 \frac{\omega_{\bm{K}}N_{\mathrm{rep}}}{2c}\ell\Bigl(\frac{2\pi}{N_{\mathrm{rep}}}\Bigr)=\pi L,\quad L=1,\dots,N_{\mathrm{rep}}-1,
 \label{bi8}
\end{equation}
and, for $N_{\mathrm{rep}}>2$, it has minor maxima if
\begin{equation}
 \frac{\omega_{\bm{K}}N_{\mathrm{rep}}}{2c}\ell\Bigl(\frac{2\pi}{N_{\mathrm{rep}}}\Bigr)=\pi L
 +\frac{\pi}{2},\quad L=1,\dots,N_{\mathrm{rep}}-2.
 \label{bi9}
\end{equation}
This pattern is exactly observed in our numerical analysis and is very well-known for the angular distribution of radiation passing through the diffraction grating. Moreover, for laser pulses considered, the global phase is a linear function of the frequency of emitted radiation and for the peak frequencies $\omega_{\bm{K},L}$ we obtain,
\begin{equation}
\Phi_{\mathrm{Th},\sigma}(\omega_{\bm{K},L})=\Bigl(N_{\mathrm{rep}}\mp\frac{1}{2}\Bigr)\pi+N_{\mathrm{rep}}L\pi.
\label{bi14}
\end{equation}
Hence, up to the same constant term, the phase is 0 modulo $\pi$, which proves the coherent properties of the Thomson combs. 
Moreover, the peak frequencies $\omega_{\bm{K},L}$ are equally separated from each other.

This pattern repeats itself for the Compton scattering, although  for the quantum process the peak frequencies $\omega_{\bm{K},L}$ are not equally separated from each other. Here, the corresponding diffraction formula adopts the form~\cite{Krajewska2014c}
\begin{equation}
\mathcal{A}_{\mathrm{C},\sigma}(\omega_{\bm{K}},\lambda_{\mathrm{i}},\lambda_{\mathrm{f}})=\exp\Bigl[\mathrm{i}\Phi_{\mathrm{C},\sigma}(\omega_{\bm{K}},\lambda_{\mathrm{i}},\lambda_{\mathrm{f}})\Bigr]
\frac{\sin(\pi\bar{Q}^+/k^0)}{\sin(\pi\bar{Q}^+/k^0N_{\mathrm{rep}})}|\mathcal{A}^{(1)}_{\mathrm{C},\sigma}(\omega_{\bm{K}},\lambda_{\mathrm{i}},\lambda_{\mathrm{f}})|,
\label{ci17}
\end{equation}
where $\mathcal{A}^{(1)}_{\mathrm{C},\sigma}(\omega_{\bm{K}},\lambda_{\mathrm{i}},\lambda_{\mathrm{f}})$ is the Compton amplitude for a single pulse and $\Phi_{\mathrm{C},\sigma}(\omega_{\bm{K}},\lambda_{\mathrm{i}},\lambda_{\mathrm{f}})$ is the Compton global phase. In the above equation $\bar{Q}^+=\bar{p}_{\mathrm{i}}^+-\bar{p}_{\mathrm{f}}^+-K^+$. For frequencies of emitted photons, $\omega_{\bm{K},L}$ with integer $L$, that satisfy the condition
\begin{equation}
\pi\bar{Q}^+/k^0N_{\mathrm{rep}}=-\pi L,
\label{ci18}
\end{equation}
we have the coherent enhancement of the Compton amplitude, which again leads to the quadratic, $N_{\mathrm{rep}}^2$, enhancement of probability distributions. However, contrary to the Thomson case, these frequencies are not \textit{exactly} equally separated from each other on the whole interval of allowed frequencies, i.e. $[0,\omega_{\mathrm{cut}}]$, where $\omega_{\mathrm{cut}}$ is the maximum frequency of generated photons in the Compton process \cite{Krajewska2014b}. When $\omega_{\bm{K}}$ approaches the cut-off value $\omega_{\mathrm{cut}}$ the spectrum of $\omega_{\bm{K},L}$ becomes increasingly denser. This means that one can get the frequency comb for Compton scattering with approximately equally spaced peak frequencies, only on some limited frequency intervals. The Compton global phase equals
\begin{equation}
\Phi_{\mathrm{C},\sigma}(\omega_{\bm{K}},\lambda_{\mathrm{i}},\lambda_{\mathrm{f}})=
-\pi\frac{\bar{Q}^+}{k^0}+\Phi_{\mathrm{C},\sigma}^{\mathrm{dyn}}(\omega_{\bm{K}},\lambda_{\mathrm{i}},\lambda_{\mathrm{f}}),
\label{ci20}
\end{equation}
where $\Phi_{\mathrm{C},\sigma}^{\mathrm{dyn}}$ is the dynamic phase \cite{Krajewska2014c}. For arbitrary laser pulses and polarizations of emitted photons the dynamic phase can only be calculated numerically. We have checked that for laser pulses considered in this paper the dynamic phase is independent of $\omega_{\bm{K}}$. Hence, for the peak frequencies $\omega_{\bm{K},L}$, the global phase, 
\begin{equation}
\Phi_{\mathrm{C},\sigma}(\omega_{\bm{K},L},\lambda_{\mathrm{i}},\lambda_{\mathrm{f}})=
\pi LN_{\mathrm{rep}}+\Phi_{\mathrm{C},\sigma}^{\mathrm{dyn}}(\omega_{\bm{K},L},\lambda_{\mathrm{i}},\lambda_{\mathrm{f}}),
\label{ci21}
\end{equation}
is the same modulo $\pi$. This does not mean, however, that the Compton frequency comb, contrary to the Thomson one, is perfectly 
coherent. This time, the distance between the peaks change a little bit, due to the recoil of electrons during the emission of photons. For the low-frequency part of the frequency spectrum these effects are rather small, but for the high-frequency part they become significant.

%-------------
   \begin{figure}
   \begin{center}
   \begin{tabular}{c}
   \includegraphics[width=13cm]{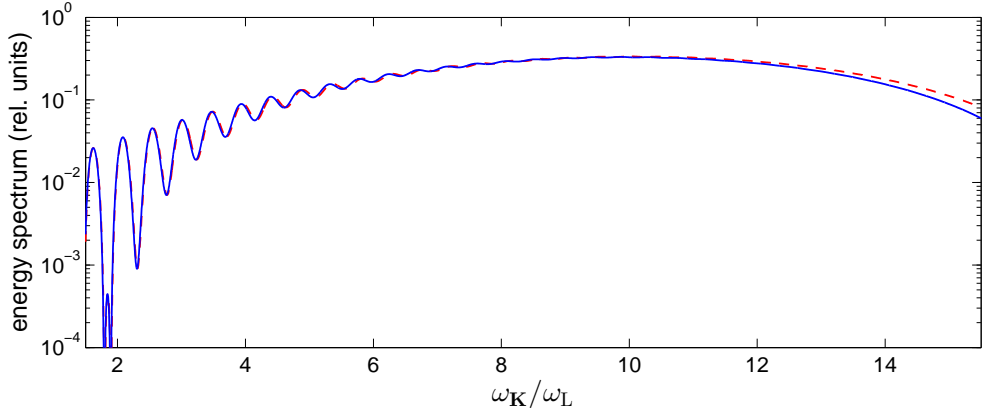}
   \end{tabular}
   \end{center}
   \caption
		{ \label{tcomb2} 
		Compton (blue line) and Thomson (dashed red line) energy distributions as a functions of frequency $\omega_{\bm{K}}$. The laser beam, linearly polarized in the $x$-direction, propagates in the $z$-direction and collides with the electron beam in the head-on geometry. The distribution is calculated in the reference frame of electrons with the laser pulse parameters such that $\omega_{\mathrm{L}}=3\times 10^{-2}m_{\mathrm{e}}c^2$, $\mu=10$, $N_{\mathrm{osc}}=3$, $N_{\mathrm{rep}}=1$. The emitted radiation is calculated for $\theta_{\bm{K}}=0.1\pi$ and $\varphi_{\bm{K}}=\pi$.
   }
   \end{figure} 
   \begin{figure}
   \begin{center}
   \begin{tabular}{c}
   \includegraphics[width=13cm]{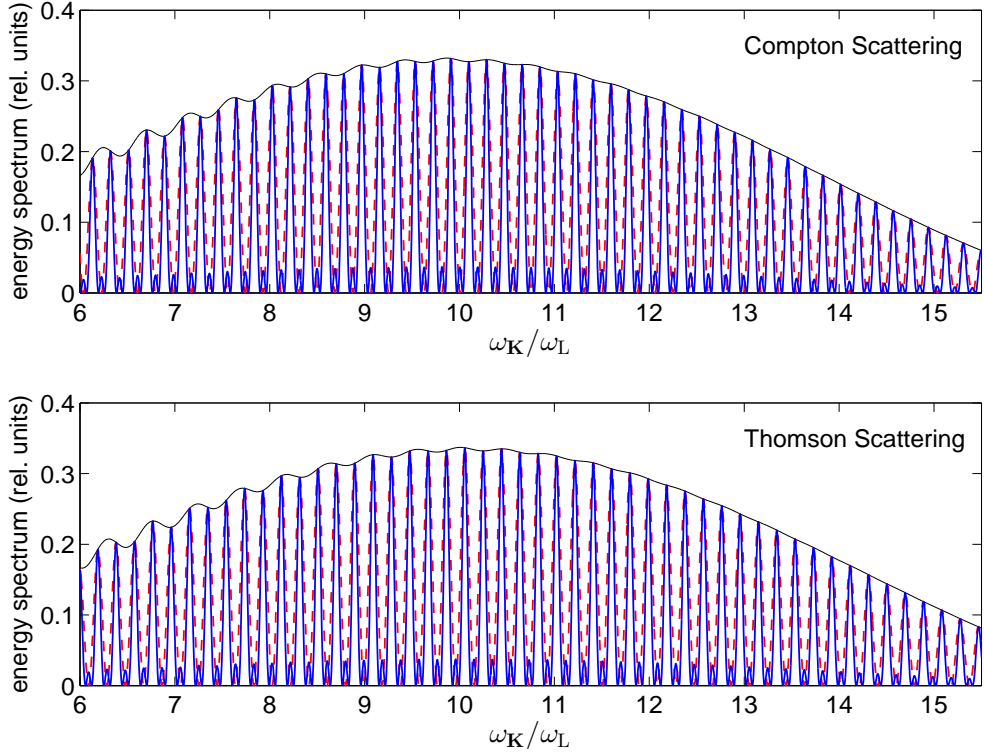}
   \end{tabular}
   \end{center}
   \caption
		{ \label{tcomb3a} 
		The same as in Fig.~\ref{tcomb2}, but for the modulated laser pulse. The thin black line (the envelope) corresponds to $N_{\mathrm{rep}}=1$, the thick dashed red line to $N_{\mathrm{rep}}=2$, the thick blue line to $N_{\mathrm{rep}}=3$, and the distributions are divided by $N_{\mathrm{rep}}^2$.
   }
   \end{figure} 
%-------------

In Fig.~\ref{tcomb1} we have presented the formation of frequency combs for relatively long laser pulses. For such pulses the combs do not have the `generic' form of distribution spreading over a broad frequency region, which for the high-order harmonic generation is called the plateau. For very short laser pulses the situation is different. Instead of well-visible and separated from each other individual peaks, we detect now broad coherent structures~\cite{Krajewska2014b} extending even to a few MeV. In our recent papers we demonstrated that, within such broad structures, it is possible to create coherent frequency combs for both the electromagnetic and the matter waves~\cite{Krajewska2014a,Krajewska2014c} by applying a finite and modulated laser pulse. This is illustrated in Fig.~\ref{tcomb2} for a single pulse, and in Fig.~\ref{tcomb3a} for a sequence of up to three such pulses. In our numerical examples the laser field parameters and the scattering geometry has been chosen such that the classical and quantum approaches display nearly identical results; for the displayed frequency region there are 50 peaks for the Compton scattering and 49 for the Thomson one, but for larger frequencies the difference becomes more significant. Similarities and discrepancies between the nonlinear Thomson and Compton processes are discussed in Ref. 8 and 16.

\section{Temporal Power Distributions}
\label{Pulse}

In order to investigate further properties of frequency combs let us consider the temporal power distribution of emitted radiation. This power distribution is related to the Compton amplitude by the formula~\cite{Krajewska2014b}
\begin{equation}
\frac{\mathrm{d}^2P_{\mathrm{C},\sigma}(\phi_{\mathrm{r}},\lambda_{\rm i},\lambda_{\rm f})}{\mathrm{d}^2\Omega_{\bm{K}}}=\frac{\alpha}{\pi}\bigl(\mathrm{Re} \tilde{\mathcal{A}}^{(+)}_{\mathrm{C},\sigma}(\phi_{\mathrm{r}},\lambda_{\rm i},\lambda_{\rm f})\bigr)^2 ,
\label{tt9}
\end{equation}
where
\begin{equation}
\mathcal{A}_{\mathrm{C},\sigma}(\omega,\lambda_{\rm i},\lambda_{\rm f})\mathrm{e}^{-\mathrm{i}\omega\phi_{\mathrm{r}}/m_{\mathrm{e}}c^2 }.
\label{tt6}
\end{equation}
Here, `$\mathrm{Re}$' denotes the real part and $\phi_{\mathrm{r}}=m_{\mathrm{e}}c^2(t-R/c)$, with $R$ being a distance from the scattering region to the observation point.

%-------------
   \begin{figure}
   \begin{center}
   \begin{tabular}{c}
   \includegraphics[width=11cm]{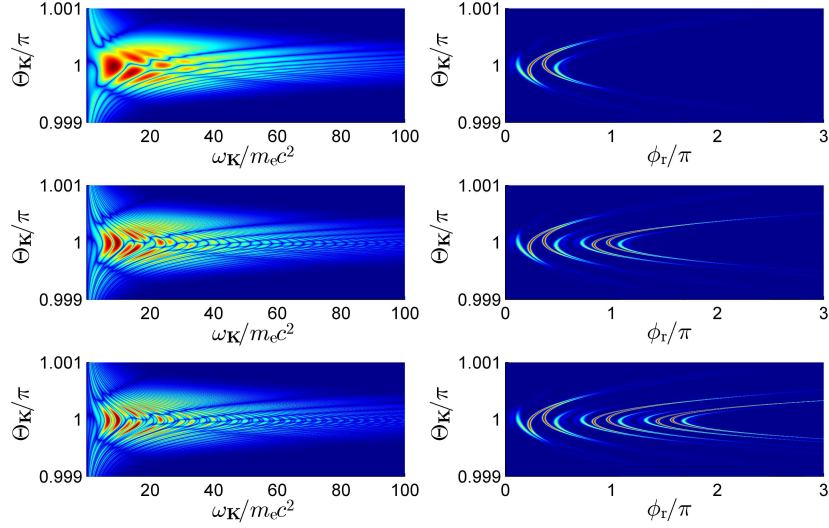}
   \end{tabular}
   \end{center}
   \caption
		{ \label{comb3r600med} 
Color mappings of the Thomson energy distribution produced in a head-on geometry of a laser beam (propagating in the $z$-direction) and an electron beam. The electric field of a driving pulse, linearly polarized in the $x$-direction, is described by the shape function with $N_{\mathrm{osc}}=3$, and $N_{\mathrm{rep}}=1$ (upper left panel), $N_{\mathrm{rep}}=2$ (middle left panel), $N_{\mathrm{rep}}=3$ (lower left panel). Its central frequency in the laboratory frame equals 
$\omega_{\mathrm{L}}=1.548\mathrm{eV}\approx 3.03\times 10^{-6}m_{\mathrm{e}}c^2$ and $\mu=1$. 
Electrons move with momentum $\bm{p}_{\mathrm{i}}=-1000m_{\mathrm{e}}c\,{\bm e}_z$ and the scattering process occurs in the $(x,y)$-plane. In the right panels we present the corresponding color mappings of the temporal power distribution of emitted radiation.	
   }
   \end{figure} 
%-------------

It is worth noting that the temporal power distribution of radiation generated by the Compton process depends in general on the electron's spin degrees of freedom. However, for the laser field parameters considered in this report the spin-flip process, for which $\lambda_{\rm i}\lambda_{\rm f}=-1$, occurs with a marginal probability, and the similarity between the quantum and classical approaches is present only for cases with $\lambda_{\rm i}\lambda_{\rm f}=1$, i.e., when the spin degrees of freedom do not change during the scattering event. The corresponding temporal power distribution of radiation generated by the classical Thomson scattering adopts the same form with the Compton amplitude replaced by the Thomson one. Because of the similarity of quantum (with no spin-flip) and classical amplitudes we shall limit our further discussion only to the classical case.

%-------------
   \begin{figure}
   \begin{center}
   \begin{tabular}{c}
   \includegraphics[width=11cm]{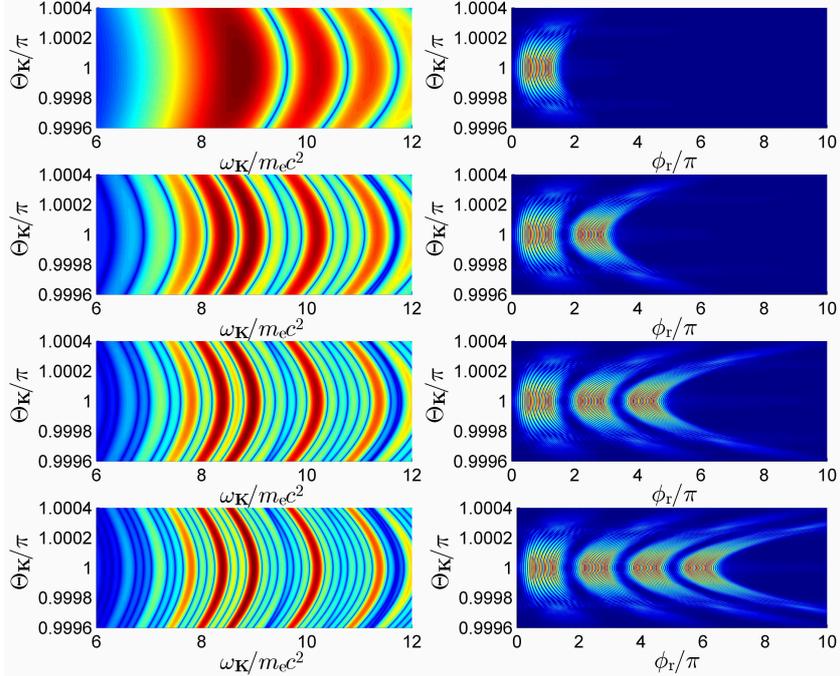}
   \end{tabular}
   \end{center}
   \caption
		{ \label{comb17r600low} 
		The same as in Fig~\ref{comb3r600med}, but for $N_{\mathrm{osc}}=17$ and $N_{\mathrm{rep}}=1,2,3$ and 4 (going from the top to the bottom).
   }
   \end{figure} 
%-------------

Our aim is now to analyze the angular-frequency and angular-time distributions of generated radiation in the Thomson scattering by modulated laser pulses. Since the Thomson and Compton processes are relativistically covariant, therefore, our general discussion presented above was in the reference frame of incident electrons. The purpose of this section is to present results in the laboratory frame. To this end we consider the pulse generated by the Ti:Sapphire laser of the central frequency $\omega_{\mathrm{L}}=1.548\mathrm{eV}$ and the electron beam of energy around $500\mathrm{MeV}$. For the head-on collision of such two beams the radiation is generated in a very narrow cone in the direction of electron beam. For this reason, in our numerical illustrations we use, instead of the spherical angles $(\theta,\varphi)$, the angular parameters $(\Theta,\Phi)$, $0\leqslant \Theta<2\pi$ and $0\leqslant \Phi<\pi$, such that~\cite{Krajewska2012}
\begin{equation}
(\theta,\varphi)=\begin{cases}
(\Theta,\Phi), & \mathrm{for}\quad 0\leqslant\Theta\leqslant\pi,
\cr
(2\pi-\Theta,\Phi+\pi), & \mathrm{for}\quad \pi<\Theta<2\pi.
\end{cases}
\label{newangles}
\end{equation}
We consider the angular-frequency and angular-time distributions for the geometry in which all momenta and polarization vectors are in the $(xy)$-plane defined by $\Phi=0$. In Fig.~\ref{comb3r600med} we present the color maps for the angular-frequency (left column) and the angular-time (right column) distributions for three laser pulses consisting of $N_{\mathrm{rep}}=1$, 2 and 3 subpulses. We see the formation of a supercontinuum in the MeV domain of frequencies which can be synthesized to very short zeptosecond pulses of radiation. It is important to note that synthesis of the Thomson energy distribution into a sequence of well-separated and ultra-short pulses of generated radiation is only possible if the phase of the Thomson amplitude is well 
approximated by the linear dependence on the frequency of emitted radiation. In fact, any significant deviation from such a rule washes away the ultra-short structure of generated radiation. It appears that the genuine quantum recoil of electrons during the emission of photons generates a nonlinear dependence of phases of the Compton amplitudes on $\omega_{\bm K}$ and, hence, it can lead to the disappearance of ultra-short temporal structures of emitted radiation for sufficiently intense laser pulses.

\section{Conclusions}
\label{Conclusions}

The existence of a broad bandwidth radiation (spanning a few MeV), which is sharply elongated around the propagation direction of the electron beam, has been demonstrated. Our analysis of temporal distributions of the observed radiation shows that it can be used for the synthesis of zeptosecond (likely even yoctosecond) pulses. Note that this is possible provided that the broad bandwidth radiation is coherent, which clearly proves that nonlinear Thomson or Compton scattering can lead to a generation of a supercontinuum. We demonstrated an important role of the global phase for synthesis of ultra-short pulse generation. Specifically, we showed that the global phase of Thomson amplitude is a linear function of the energy of emitted radiation, $\omega_{\bm K}$. This guarantees that emitted radiation is coherent and can be synthesized into ultra-short pulses. In addition, we investigated a possibility of generating coherent frequency combs from Thomson and Compton scattering in the presence of a sequence of short subpulses. This was motivated by the celebrated high-order harmonic generation and by the resulting synthesis of attosecond pulses out of the frequency spectrum of those harmonics combs. We studied here the generation of frequency-comb structures for the ideal situation when all subpulses are identical. Such a situation can be well-modeled by composing laser pulses from a few monochromatic ones. In fact, the laser pulse shapes considered in this paper are composed either of three monochromatic components with appropriately chosen amplitudes ($N_{\mathrm{rep}}>2$) or only of two for $N_{\mathrm{osc}}=2$. This fact raises the question: How sensitive is the formation of frequency combs if we change relative phases of these monochromatic components? This and similar problems are currently investigated.

%%%%%%%%%%%%%%%%%%%%%%%%%%%%%%%%%%%%%%%%%%%%%%%%%%%%%%%%%%%%%
%%%%% References %%%%%

\end{document}